\documentstyle[12pt]{article}

\begin{document}

\hoffset = -1truecm
\voffset = -2truecm

\title {\bf 2D Supergravity and Integrable Systems} \\
\author {  {\bf R. Ch. Rashkov } \\ \\
 International Centre for Theoretical Physics, Trieste 34100,
{\bf Italy}\thanks{Permanent Address: Department of Physics, Sofia
University, 5 J.Baucher Blvd., 1126 Sofia, Bulgaria } \\ \\ \\ \\ \\
\date{30th October 1992}}

\maketitle

\begin{abstract} \\

In this letter we consider a realization of Osp (2$\mid$2) Kac-Moody algebra
and its connection with the integrable systems. Integrable hierarchy based on
the constrained Osp (2$\mid$2) connection is constructed. Using zero-curvature
formulation we elucidate the connection between the integrable hierarchy and 2D
supergravity. It is shown that super-Virasoro transformations are symmetries of
the our hierarchy. The analogies with W$_3^{(2)}$ algebra are given.
\end{abstract}

\newpage

\section{Introduction}

In the last years it becomes that the interest in physics of integrable
hierarchies increase more and more. It was recognize a deep connection between
2D supergravities and integrable hierarchies in the matrix model formulation
and  the continuous theory as well.

The most natural approach to derive the bi-Hamiltonian structure of the system
under consideration  is so called zero-curvature formulation. In this
formulation the time evolution of the system can be thought of as
integrability condition (zero-curvature) associated with some gauge group. In
our case this is Osp (2$\mid$2) current algebra. So called Drinfeld-Sokolov
scheme provide a procedure, which by constraining the current algebra one
obtain
the KdV hierarchies with as gauge group the Virasoro algebra.

In 2D gravity such constrains are proposed in [1] and the used procedure is
referred to also as "Polyakov soldering" procedure. In [1,2] was proposed also
a new soldering in the case of Sl$_3$ current algebra, which lead to so called
W$^{(2)}_3$ algebra. This algebra provide properties closely similar to N=2
super-Virasoro algebra. Using "Polyakov  soldering" procedure for Osp
(2$\mid$2) supersymmetric Wess-Zumino model [3,4] it is straightforward  to
derive super-Virasoro algebra as symmetry of the constrained system (which is
in fact 2D supergravity). So that the formulae in the case of
Polyakov-Bershadsky soldering of Sl$_3$ and the usual supersymmetric case can
be compared. The behaviour of the fields in both cases are similar. This gives
us a  reason to treate such algebras together, namely as quasi-super algebras.

In the approach, related to integrable systems, the
case of W$^{(2)}_3$ algebra was elaborated by van Driel [5]. In this direction
using algebraic approach [6], referred to as Wilson's dressing method [7] it
is interesting to elaborate the supersymmetric case and to compare the results
to those of van Driel. It seems that extension to the general case is more or
less straightforward and the example, treated in this letter can be considered
as a step in this direction.

In [5] a twisted
embedding Sl$_2 \subset Sl_3$ was used in order to consider twisted realization
of A$^{(1)}_1$ which exist beside the principal and  homogenious ones. This
naturally lead to W$^{(2)}_3$ algebra. The above realization is connected to
corresponding realization of the Heisenberg algebra, which later is used to
define of time evolution operator H.

In this letter our aim is to study zero-curvature formulation associated with
Osp (2$\mid$2) current superalgebra as a gauge group. This will allowed us to
compare  our results to these of  [3] and to make the relation with 2D
supergravity.	The second step is to prove that super-Virasoro algebra
transformations are symmetries of the corresponding hierarchy. Considering
super-$\tau$ function  as the group orbit of the highest weight vector of Osp
(2$\mid$2) representation an explicit form of the basic fields in terms of
$\tau$-function will be given.
\ \\
\ \\

\section{2D supergravity and integrable systems}

Let us remember some features of the theory of 2D supergravity and its
connection to supersymmetric WZ model. It is well known that the Hamiltonian
reduction of Osp (2$\mid$2) supersymmetric WZ theory with respect to maximal
parabolic subgroup lead to super-Virasoro algebra. The most direct approach to
find it is so called "Polyakov soldering" in which the Osp (2$\mid$2)
constrained connection is of the form:
\begin{equation}
A = \left(
\begin{array}{ccc}
\Phi & T & \zeta \\
1 &  \Phi & 0 \\
0 & \eta & 2\Phi \end{array} \right)
\end{equation}
\noindent It is straightforward to use the transformation law for the above
constrained
form of the connection in order to derive the transformation properties of the
matrix elements $T$, $\Phi$, $\zeta$, $\eta$ [4]. Namely:
$$ \delta A = \partial  \Lambda + [\Lambda, A] $$
\noindent where $\Lambda \in$ Osp (2$\mid$2). In components we have [9]
equations four of
which are independent [4]. The substitution  of the super-Laurent expansion of
the fields in the above equation give us the familiar N=2  super-Virasoro
commutation relations:
$$ \left[ L_n, L_m\right] = (n-m) L_{n+m} + {n(n^2-1)\over 8} \delta_{n+m}, $$
\begin{equation}
 \left[L_n, G_{\alpha}\right]  = \left(n - {1\over 2} \alpha\right) G_{\alpha +
n}
\end{equation}
$$ \left\{G_{\alpha}, G_{\beta} \right\} = {1\over 2} L_{\alpha + \beta} +
{1\over
8} \left(\alpha^2 - {1\over 4} \right) \delta_{\alpha + \beta} $$

So, it can be constituted that the super-Virasoro algebra is the symmetry of
our
Osp (2$\mid$2) WZ theory and therefore it is equivalent to 2D supergravity.

It is straightforward procedure to derive the corresponding Ward identities and
operator-product expansion [4].

In this section we will formulate the zero-curvature relation referred to as
Wilson's dressing method [6,7]. At first, we will consider Osp (2$\mid$2)
algebra by the generators:
\begin{equation}
\left\{ e, f, h, h_o, \varepsilon^+, \varepsilon^-, \varphi^+,
\varphi^-\right\} = \left\{ t_i, t_{\alpha} \right\}
\end{equation}
\noindent fulfilling the commutation relations:
\begin{equation}
\left[ t_A, t_B \right. \left. \right\} = f^C_{AB} t_C, \ \ \ \  A,B,C =
\left\{i, \alpha \right\},
\end{equation}
\noindent where $f^C_{AB}$ denote the structure constants of the algebra.

The subject of our consideration will be Kac-Moody extension of the Osp
(2$\mid$2) superalgebra and its representation	$L(\Lambda_o)$ with highest
vector $\Lambda_o$. Our aim is to build up algebraic picture of the
zero-curvature formulation based on the algebra $\hat g$ consists of
elements of the form:
\begin{equation}
M = \left( \begin{array}{cc} M_1 & F_1 \\ F_2 & M_2 \end{array} \right), str M
= tr M_1 - tr M_2
\end{equation}
\noindent $M_1$ and $M_2$ are bosonic while $F_i$ are fermionic.

The superalgebra $\hat g$ contains an infinite-dimensional Heisenberg
subalgebra generated by the system $\{p_i, q_i\}$ satisfying the following
commutation relations:
\begin{equation}
[p_i, q_j] = k \delta_{ij}
\end{equation}

One can choose the following realization of $p_i$ and $q_j$:
\begin{equation}
p_1 = \lambda e + f + \lambda^{1/2} \varepsilon^+ + \varphi^+
\end{equation}
\begin{equation}
q_1 = e + \lambda^{-1} f + \lambda^{-1/2} \varphi^- + \varepsilon^-
\end{equation}
\begin{equation}
p_i = p^i_1
\end{equation}
\begin{equation}
q_i = \mid \omega (p_i) \mid
\end{equation}
\noindent where $i$ is odd and $\omega$ is antilinear involution used later to
define
contravariant hermitian form. Our aim is using the above Heisenberg algebra to
construct Lax connection, after which to study the zero-curvature formulation.

In order to realize the above scheme, following [7] we have to introduce the
time evolution operator for our system:
\begin{equation}
H = \sum^{\infty}_{i=1} p_i t_i
\end{equation}

Let us suppose that we have highest weight module $L(\Lambda_o)$, on which we
have defined an action of operator of the translations $T^l$. Let us also
define the following loop group element:
$$
\hat{\psi}^l = T^{-l} e^{H} g
$$

Based on Mulase [8] in	our case case the Birkhoff decomposition is hold and
therefore:
\begin{equation}
\hat{\psi}^l = \hat{\psi}_+ \hat{\psi}_o \hat{\psi}_-
\end{equation}
\noindent where every subgroup consists in even and odd part.

In addition:
\begin{equation}
\partial_{t_i} \hat{\psi}^l = p_i \hat{\psi}^l
\end{equation}
\noindent or:
\begin{equation}
\nabla_i = \partial_{t_i} - \left[ \left(\hat{\psi}^L_- \right)^{-1} p_i
\psi_-\right]_+
\end{equation}
\noindent where "+" denotes the positive powers of $\lambda$ ($\lambda$ is the
spectral
parameter). Considering the loop group processing Birkhoff decomposition we
have that every element at the loop group level admits a factorization [7,12]:
$$
M = M_+ M_o M_-
$$
\noindent where
\begin{equation}
M_- = 1 + \theta \lambda^{-1/2} M_{-1/2} + \lambda^{-1} M_{-1} + ...
\end{equation}
$$
M_+ = 1 + \theta \lambda^{1/2} M_{1/2} + \lambda^{1} M_{1} + ...
$$
\noindent and $M_{\pm A}$ are arbitrary elements of Osp (2$\mid$2) and
\begin{equation}
H_o = N_- w H N_+,
\end{equation}
\noindent $N_{+(-)}$ is  upper (lower) triangular, $w$ - element of super Weyl
group, $H$
- Cartan matrix.

Using the loop group decomposition and the formula (14) for the connection $A$,
it immediately follows that:
\begin{equation}
\left[\left(\hat{\psi}_-\right)^{-1} p_i \hat{\psi}_-\right]_+ = A = \left(
\begin{array}{ccc}
n & p+\lambda & \alpha + \theta \lambda \\
1 & m & \beta \\
\delta & \gamma - \theta \lambda & n + m \end{array}
\right)
\end{equation}
\noindent and our covariant derivative looks like:
\begin{equation}
\nabla = \partial_{t_i}  - \left( \begin{array}{ccc}
n & p+\lambda & \alpha + \theta \lambda \\
1 & m & \beta \\
\delta & \gamma - \theta \lambda & n + m \end{array}
\right)
\end{equation}

But it is known that the above connection is invariant under the Borel
subgroup, generated by matrices  of the form:
\begin{equation}
B = \left( \begin{array}{ccc}
1 & a & b \\
0 & 1 & 0 \\
0 & c & 1 \end{array}\right) \hbox{\hspace{0.5cm}} B^{-1} =
\left( \begin{array}{ccc} 1 & bc - a & -b \\
0 & 1 & 0 \\
0 & -c & 1 \end{array} \right)
\end{equation}
\noindent In order to put $A$ in the form (1) using Borel subgroup, generated
by (19) we
perform gauge transformations of the form:
\begin{equation}
A \to A' = \partial B B^{-1} + B^{-1} AB
\end{equation}
\noindent The elements of the matrix $A$ takes the form (1) where we have used:
$$
A = \left( \begin{array}{ccc}
n & p & \alpha \\
q & n & -\beta \\
\delta & \gamma & n+m \end{array} \right)
$$

$$
\Phi = {n+m + \beta \delta \over 2}
$$
\begin{equation}
\zeta = \alpha - 2 \beta \Phi - n \beta - \partial \beta
\end{equation}
$$
\eta = \gamma + 2 \delta \Phi - m \delta + \partial \delta
$$
$$
T = p + \alpha \delta + \beta \gamma + 2 \Phi^2 - 2 n \Phi - \beta \delta n -
(\delta \beta)^2 + \partial (n - \Phi) + \partial \beta \delta - 2 n m - n^2
$$
\noindent Two remarks are in order. First one is that we have considered a
gauged
supersymmetric WZ model possessing Osp (2$\mid$2) symmetry. Using the
zero-curvature formulation we have derive  constrained connection $A$.

The second is that using the above connection it can be derived the
super-Virasoro algebra as a symmetry of the reduced model.

We note that the transformation properties can be extracted from the
zero-curvature formulation. Let us denote $A = J_x$. From the integrability
conditions it follows that:
\begin{equation}
{\partial J_x \over{ \partial t_i}} = \left[ J_i, J_x \right] + J'_i
\end{equation}
\noindent where $J_x$ is the constrained connection but $J_i$ is in general
form:
\begin{equation}
J_i = \left( \begin{array}{ccc}
n & p & x \\
q & m & \beta\\
\delta & \gamma & n + m \end{array} \right)
\end{equation}
\noindent Then
\begin{equation}
{\partial J_x\over {\partial t_i}} =  \left( \begin{array}{ccc}
p - T q - \zeta \delta + n' & (n-m) T + \alpha \eta - \zeta \eta + p' &
\Phi \alpha - T \beta - \zeta m + \alpha '\\
n - m + q' & q T + \beta \eta - p + m' & q \zeta - \alpha + \beta \Phi + \beta
'\\
\gamma - q \eta + \delta \Phi + \delta ' & \delta T + n \eta - \Phi \gamma +
\gamma ' & \beta \eta - \zeta \delta + n' + m'
\end{array} \right)
\end{equation}
\noindent Because of the constrained from of $J_x$, from the above equation we
can solve
the parameters $m$, $\gamma$, $\alpha$ in terms of the other:
$$
m = n + q'
$$
$$
\gamma = q \eta + \delta \Phi - \delta '
$$
\begin{equation}
\alpha = q \zeta + \beta \Phi + \beta '
\end{equation}
$$
p = T q + {\zeta \over 2} \delta - {\beta \over 2} \eta + {q''\over 2}
$$
\noindent Then simple calculations allow us to derive the equations of motion
for the
fields $T$, $\zeta$, $\eta$ and $\Phi$ (in  $t_i$ direction)
$$
{d \Phi \over {d t_i}} = - {\eta \beta + J \delta \over 2} +
\left( {n + m \over 2}\right)'
$$
$$
{d \zeta\over {d t_i}} = \left[ (\partial + \Phi)^2 - T \right] \beta - \zeta m
+ \left[ \zeta \partial + \zeta ' + \Phi \zeta \right] q
$$
\begin{equation}
{d \eta \over { d t_i}} = - \left[ (\partial - \Phi)^2 - T \right] \beta + \eta
n + \left[ \eta \partial + \eta ' - \Phi \eta \right] q
\end{equation}

$$
{d T\over {d t_i}} = \left[ {1\over 2} \partial^3 + 2 T \partial + \partial T
\right] q + \left[ {3\over 2} \zeta \partial + {1\over 2} \partial \zeta - \Phi
\zeta \right] \delta - \left[ {3\over 2} \eta \partial + {1\over 2} \partial
\eta + \Phi \eta \right] \beta
$$

The following remarks are in order:

1. The above equations of motion coincide with those of W$^{(2)}_3$ algebra
found in [5]. The only difference is that the fields $\zeta$ and $\eta$ in our
case are fermionic  while in [5] they are bosonic.

2. Equations (26) determine Poisson structures.
\ \\
\ \\

\section {Super-Virasoro transformations as gauge symmetries of the
supersymmetric hierarchy}

We have to extend the results of [9] to the supersymmetric case. For this
purpose let us consider $A$ as a component of the Lax pair connection $[-P_k,
A]$. The super-Virasoro algebra is in fact the algebra of gauge transformations
preserving the form of the Lax connection. Remember the transformation rules
(22) and the constrained form of $A$ it can be checked that $\Lambda$ is fixed
to be:
\begin{equation}
\Lambda = \left( \begin{array}{ccc}
n & {1\over 2} q'' + Tq + {1\over 2} (\zeta \delta + \eta \beta) & q \zeta +
\beta \Phi + \beta ' \\
q & n + q' & \beta \\
\delta & q \eta + \delta \Phi - \delta ' & 2 n + q' \end{array} \right)
\end{equation}
\noindent where $n, q, \delta , \beta$ are parameters and the transformation
properties
of $T, \zeta, \eta$ and  $\Phi$ to be as in eqs. (26). This is nothing but
super-Virasoro algebra. In the same way we can consider arbitrary $t_k$. In
this case we will denote:
$$
\Lambda_k = \left( \begin{array}{ccc}
n_k & p_k & \alpha_k \\
q_k & n_k & \beta_k \\
\delta_k & \gamma_k & 2 n_k \end{array} \right)
$$
\noindent and the integrability condition reads off:
\begin{equation}
\left[ - \partial_k + \Lambda_k, \partial + \Lambda \right] = o
\end{equation}
\noindent where:
\begin{equation}
\Lambda_k = \left( \begin{array}{ccc}
n_k & {1\over 2} q_k'' + Tq_k + {1\over 2} (\zeta \delta_k + \eta \beta_k) &
q_k \zeta + \beta_k \Phi + \beta_k ' \\
q_k & n_k + q_k' & \beta_k \\
\delta_k & q_k \eta + \delta_k \Phi - \delta_k ' & 2 n_k + q_k' \end{array}
\right)
\end{equation}
\noindent Then, using the transformation properties of $\Lambda_k$:
\begin{equation}
\delta \Lambda_k = \partial_{t_k} \Lambda + \left[ \Lambda_k, \Lambda \right]
\end{equation}
\noindent it is easy to derive the conditions which must be satisfied by $q,
\delta,
\beta$ and $n$ the form of $\Lambda_k$ to be preserved. Therefore they must
satisfy:
\begin{equation}
\partial_{t_k} q = \hat{\delta} q_k + q'_k q - q'q_k + \beta_k \delta - \beta
\delta_k
\end{equation}
\begin{equation}
\partial_{t_k} \delta = \hat{\delta} \delta_k + n_k \delta + n\delta_k + \Phi
\delta q_k - \Phi \delta_k q + \delta '_k q - \delta q'_k
\end{equation}
\begin{equation}
\partial_{t_k} \beta = \hat{\delta} \beta_k - n_k \beta - n\beta_k + \beta_k
\Phi q - \beta \Phi q_k + \beta '_k q - \beta ' q_k
\end{equation}
\begin{equation}
\partial_{t_k} n = \hat{\delta} n_k + {1\over 2} (\zeta \delta + \beta \eta)
q_k - {1\over 2} (\zeta \delta_k + \beta \eta_k) q + (q \zeta + \beta \Phi +
\beta ') \delta_k - (q_k \zeta + \beta_k \Phi + \beta_k') \delta
\end{equation}
\noindent where $\hat{\delta}$ denote variation of the corresponding
generalized
Gelfand-Dickii polinomial.

In the rest we will give an explicit expression of the fields $T, \eta$ and
$\zeta$ in terms of $\tau$-function. The key point of our consideration is the
representation of $\tau$-function as "vacuum expectation value":
\begin{equation}
\tau (t) = <v_{\Lambda_o}, \psi (t) g v_{\Lambda_o} >
\end{equation}
\noindent and decomposition:
\begin{equation}
\tau (t) = \sum \tau_l \bigotimes W_l
\end{equation}
\noindent where $W_l$ is  the $l^{th}$ weight vector which belong
to the module $V_{\Lambda_o}$.

We have defined also covariant derivative (14), compatibility conditions for
which leads to the zero-curvature:
\begin{equation}
\left[ \nabla_i, \nabla_j \right] = 0
\end{equation}
\noindent To be able to evaluate the components of $A$ from the   above
equations let us
consider the simplest case in which $t_1 = x$. The explicit form of the
covariant derivative reads off:
\begin{equation}
\nabla_x = \partial_x - A_x
\end{equation}
\noindent where $A_x$ is given by (1,21).

We are in a position to project out the connection $A_x$ on $h, h_{(0)}, f$ and
$\varphi$ subspaces and to give an explicit expression for $T$, $\zeta$ and
$\eta$ in terms of $\tau$-function defined as in (43,44). Omitting the detailed
derivation, which will be given elsewhere [10], we give below the final result:
\begin{equation}
\Phi = \partial \log \tau_l
\end{equation}
\begin{equation}
T = - 2 \partial^2 \log \tau_l
\end{equation}
\begin{equation}
\zeta = \theta_1 {\tau_{l + 1}\over {\tau_l}}
\end{equation}
\begin{equation}
\eta = - \theta_2 {\tau_{l - 1}\over {\tau_l}}
\end{equation}
\noindent where $l$ are even integers. Using the equations of the hierarchy one
can
prove  that:
\begin{equation}
\zeta \eta  \sim  - 2 \partial^2 \log \tau_l
\end{equation}
\noindent Ones again we note the analogy with W$^{(2)}_3$ case [5].
\ \\
\ \\

\section{Discussions and conclusions}

In the second section the Heisenberg system for the case of Osp (2$\mid$2)
current superalgebra is constructed. This allowed us to develop the
zero-curvature method to construct the constrained connection referred to as
"Lax connection". The equivalence of the above formulation with those in the
case of "Polyakov soldering" in 2D supergravity is explained. In the last
section we have considered the super-Virasoro algebra as the group of gauge
transformations of corresponding hierarchy. We would like to note that it was
recognize whole analogy to the case of W$^{(2)}_3$ algebra. Therefore,
investigation of systems possessing  so called fractional hierarchies is
equivalent to study some supersymmetric systems. Projection of the component of
the root decomposed connection give a representation of the dynamical fields in
terms of the famous $\tau$-function and the exact form is given without
systematical proof. The detailed derivation of this representation, as well as
the all omitted calculations and proofs will be given elsewhere [10].

We would like to comment some open questions. It would be interesting to give
zero-curvature formulation in the  general case of supersymmetric hierarchies
and to give a classification of the fractional hierarchies covered by them.

Another open question is the relation to the supermatrix models. Further
investigations must  make clear the link with topological models of 2D
supergravity.

In conclusion, the geometrical meaning of the above supersymmetric hierarchy
and the connection with superflag manifolds may be an important clue to the
direction of further progress.
\ \\
\ \\

\section{acknowledgements}

This work was partially supported by Grant under Contract 214 with the Ministry
of Education and Science. The author would like to thank ICTP for hospitality
during which this work was completed.
\ \\
\ \\

{\bf References}
\begin{enumerate}
\item A.M.Polyakov. Int.J.Mod.Phys.A5 (1990) 833.
\item M.Bershadsky. Princeton Preprint IASSNS-HEP-90/44.\\
F.A.Bais, S.Tjin, P.van Driel. Nucl.Phys.B 357 (1991) 632.
\item R.Ch.Rashkov. Mod.Phys.Lett. A5 (1990) 991.\\
 R.Ch.Rashkov. Mod.Phys.Lett. A5 (1990) 2385.
\item L.J.Romans. Nucl.Phys. B357 (1991) 549.
\item P.van Driel. Phys.Lett. 247B (1992) 179, and references therein.
\item M.J.Bergvelt, A.P.E.ten Kroode. J.Math.Phys. 29 (1988) 1308.
\item G.Wilson. Compt.Rend.Acad.Sci. Paris I 299 (1985) 587.
\item M.Mulase. Invent. Math. 92 (1988) 1.
\item B.Spence. Phys.Lett. 276B (1992) 311.
\item R.Ch.Rashkov, in preparation.
\end{enumerate}

\end{document}